\documentclass[12pt]{article}
\usepackage{epsfig}

\newcommand{\tdm}[1]{\mbox{\boldmath $#1$}}
\newcommand{\cor}[1]{\left\langle{#1}\right\rangle}
\newcommand{\f}[2]{\frac{#1}{#2}}

\newcommand{\eq}{\begin{equation}}
\newcommand{\eqx}{\end{equation}}
\newcommand{\eqn}{\begin{eqnarray}}
\newcommand{\eqnx}{\end{eqnarray}}

\newcommand{\lab}{\label}

\newcommand{\al}{\alpha}

\title{{\bf RECOVERING CORRECTIONS IN THE ANALYSIS OF INTERMITTENT DATA}
       \thanks{e-mail: {\tt beataz@qcd.ifj.edu.pl}}}
\author{Beata Ziaja\\
        \it Department of Theoretical Physics\\
        \it Institute of Nuclear Physics\\
        \it Radzikowskiego 152, 31-342 Cracow, Poland\\}
\date{March 1999}

\begin{document}

%
%
%
%

\maketitle

\begin{abstract}
The analysis of intermittent data is improved. The standard method of
recovering the history of a particle cascade is proved in general {\bf not}
to reproduce the structure of the true cascade. The {\bf recovering corrections}
to the standard method are proposed and tested in the framework 
of multiplicative cascading models.
\end{abstract}
\section{Introduction}

The first data on possible intermittent behaviour in multiparticle
production \cite{l1} came from the analysis of the single event of high multiplicity
recorded by the JACEE collaboration \cite{l2}. Data from some
accelerator experiments \cite{acc} confirmed afterwards that there are large
dynamical fluctuations appearing in the high energy multiparticle final
states which manifest a scaling behaviour.
Many different models have been proposed since to explain  the effect
\cite{l4}. Some of them suggested that an underlying final state multiparticle
cascade may be responsible for the scaling of multiparticle moments
\cite{l4}. In this approach the intermittent data represent the last
stage of the cascade, and the main problem lies in the extraction of the 
information on previous cascading stages which is in some way encoded in
the last stage data. It should be stressed that the problem of recovering 
the history of the cascade may not be solvable if adressed generally. However, 
the self-similar processes which are assumed to underly the final state 
structure obey some scaling law. This makes many features of local dynamics 
dissapear, and one may expect to extract from the last stage data at least 
a part of information on the real cascade parameters .    

The method of recovering the history of the cascade from the last stage data
was proposed and applied originally to the JACEE event data. Since that 
time it became a standard tool of multiparticle data analysis \cite{l4}, 
especially in the event-by-event analysis \cite{chin}.

In this paper we would like to improve the standard method of analizing the 
intermittent data, taking into account corrections due to the recovering 
the history of the particle cascade in the framework of multiplicative
random cascading models \cite{models,bsz}. This problem has been already
adressed and analized in part in Ref.\ \cite{pesch}. Our discussion will 
proceed as follows. 
In section 2 we characterize briefly the standard method of estimation
of intermittency exponents, and introduce the definition of recovering 
corrections.
In section 3 the definition of multiplicative models is presented, and the
special cases of multiplicative models: $\alpha-$, $p-$, $(p+\alpha)-$models
are summarized.
In section 4 the recursive equation for recovering corrections in
a multiparticle model with possible neighbour-node memory is derived.
Section 5 is devoted to the implementation of recovering corrections into
the analysis of data. The implementation algorithm is proposed and
numerically tested.
Finally, in section 6 we present our conclusions.

\section{Standard estimation of intermittency exponents and recovering
corrections.}

Consider a sample of $M$ bins describing an individual (intermittent) event.
For simplicity assume that $M=2^n$, where $n$ is a natural number. We thus
have $2^n$ numbers describing the content of each bin~:
\eq
x^{(n)}_i,\,\,\, i=0,1,\ldots,2^n-1
\lab{r1}
\eqx
which represent e.\ g.\ the distribution of particle density into bins.
One assumes that the bin ensemble has been generated in some cascading
process, and the {\bf unnormalized density moments} $z^{(n)}_q$ 
for this process~:
\eq
z^{(n)}_q=\f{1}{2^n} \sum_{i=0}^{2^n-1} \left( x^{(n)}_i \right)^q
\lab{r2}
\eqx
manifest a scaling behaviour parametrized by intermittency exponents 
$\phi_{q}$~:
\eq
z^{(n)}_q \sim 2^{n\cdot \phi_q}.
\label{r3}
\eqx
%

The standard method of estimation of intermittency exponents was introduced
firstly for the analysis of JACEE event \cite{l2}. The method recovered the history 
of the cascade in the following manner:
it established the value of density moments for each cascade step, and made the
linear $\chi^2-$fit to the points $(k,\log z^{(k)}_q)$
($k=1,\ldots,n$, $\log x \equiv {\rm \log }_2 x$)~:
\eq
\label{r44}
\log z^{(k)}_q = k\cdot \phi'_q + b.
\eqx
In this way the eventual long-range correlations could be separated
and would not contribute to the estimated slope $\phi_q$.
For the assumed bin-into-two-bins splitting scheme 
the true value of the $x_i^{(n-k)}$ bin content was replaced by
$y^{(n-k)}_i$~:
\eq
y^{(n-k)}_i= \f{1}{2^k} \sum_{j=0}^{2^k-1}
x^{(n)}_{2^k\times i+j}.
\lab{r5}
\eqx
The intermittency exponents were extracted from the reconstructed moments\\
$z_{q;\,rec.}^{(k)} $~:
\eq
z^{(k)}_{q;\,rec.}=\f{1}{2^k} \sum_{j=0}^{2^k-1}  \left(
y^{(k)}_j \right)^q,
\lab{r6}
\eqx
assuming their power law behaviour~:
\eq
z^{(k)}_{q;\,rec.} \sim 2^{k\cdot \phi'_q}.
\lab{r7}
\eqx
So far (see e.\ g.\ \cite{bz}) one has estimated the value of (normalized)
$\phi_q$, assuming simply that $\phi_{q;norm.}=\phi_q'$ 
($\phi_{q;\,norm.\ }:=\phi_q-q\phi_1$). 

There is, however, an open question if and how
the cascade recovered from data refers to
the true cascade which generated the data. For the purpose of estimating
the intermittency exponents it is enough to ask about the relation between
the true density moments $z_q$ and the reconstructed ones obtained from
(\ref{r6}). 
It is obvious that formula (\ref{r5}) looses a piece of information
on the primary cascade.
We give a simple example to illustrate the problem. If we assume
that the underlying cascading process preserves e.\ g.\ 
the total particle density
{\footnotesize $\sum_{j=0}^{2^k-1} x^{(k)}_j = 1$ ($k=1,\ldots,n$)},
then it follows from (\ref{r6}) 
that the reconstructed cascade will not manifest this property: 
{\footnotesize $\sum_{j=0}^{2^k-1} y^{(k)}_j \neq 1$}. It means that
standard method does not recover the conservation law present in the 
true cascade.

Moreover, it was found explicitly for the $\alpha-$model \cite{models,bsz} 
(which does not preserve the total particle density) that there exists always discrepancy
between the true value of $\phi_q$ and its estimation $\phi'_q$ (\ref{r7}),
due to recovering technique (\ref{r5}). This problem was discussed
in detail  in Ref.\ \cite{jz}.

Assume we may express the discrepancy between the true and the reconstructed 
moments at the $(n-k)$th cascade step in a following way~:
\eq
z^{(n-k)}_{q;\,rec.}=
z^{(n-k)}_q \cdot p_q(k),
\lab{r8}
\eqx
where the factor $p_q(k)$ denotes the corrections due to
recovering procedure (\ref{r5}). We will call them recovering corrections.
Corrections $p_q(k)$ contain information on the parameters of a specific
process which generated the true cascade. It is obvious that they depend
also on the cascade step.
Substituting (\ref{r8}) into (\ref{r44}) one arrives at
the relation~:
\eq
\label{r10}
\log z^{(n-k)}_{q;\,rec.} -\log(p_q(k)) = (n-k)\cdot \phi_{q;\,corr.} + b,
\eqx
where the fitted slope $\phi_{q;\,corr.}$ estimates the true intermittency
exponent $\phi_q$.
%
%

In this paper we confine ourselves to recovering corrections considered 
for the class of multiplicative random cascading models. 
For the multiplicative cascade
("multiplicative" means that at each cascade step 
the bin content is multiplied by a number to generate the bin content 
at the next cascade step) relation (\ref{r8}) holds explicitly, and recovering
corrections take the form \cite{jz} (for proof see Appendix A)~:
\eq
\label{r13}
p_q(k)=\cor{\left( \f{1}{2^k}\sum_{i=0}^{2^k-1} x^{(k)}_i \right)^q}
\eqx
where the average $\cor{\ldots}$ is taken over the (eventual) random choices
while generating the cascade. 
The starting bin $x^{(0)}_0$ is set equal 1. It is worth noticing
that $p_q(k)$ may be also expressed in terms of the erraticity 
moments \cite{hwa}~:
\eq
p_q(k)=C_{1\, ,\,q}.
\eqx
%

\section{Multiplicative models}

As already mentioned, in our paper we restrict ourselves 
to the class of multiplicative random
cascading processes with possible neighbour-node memory, generating the
uniform distribution of particle density.
The commonly used models of random cascading: $\al-$model \cite{alfa}
and $p-$model \cite{MS87}
belong to this class. For the purpose of testing our predictions for recovering
corrections we introduce below a new many parameter $(p+\al)-$model.

In the multiplicative random cascading processes with possible neighbour-node 
memory assume for simplicity the root of a cascade to be equal to 1~:
$x_0^{(0)}=1$. One generates the next stages of the cascade recursively.
The scheme is following. 
The two bins $x^{(k+1)}_{2i}$ and $x^{(k+1)}_{2i+1}$
are obtained from $x^{(k)}_i$ by multiplication~:
\eqn
x^{(k+1)}_{2i}   &:=& W_1 \cdot x^{(k)}_i,\nonumber\\
x^{(k+1)}_{2i+1} &:=& W_2 \cdot x^{(k)}_i ,
\lab{r14}
\eqnx
where $W_1$ and $W_2$ are random variables of the $m$ model parameters
$a_j$, $j=1,\ldots,m$~:
\begin{center}
$W_1=a_j$ with probability $p_{a_j}$,\\
$W_2=a_j$ with probability $p_{a_j}$,
\end{center}
\eq
\lab{r1414}
\eqx
with normalized probability weights
$p_{a_j}$~:
\eq
\sum_{j=1}^{m}p_{a_j}=1.
\lab{r15}
\eqx
The distribution of particle density will be uniform if the following
condition is fulfilled~:
\eq
p(W_1=a_i,W_2=a_j)=p(W_1=a_j,W_2=a_i),
\lab{unif}
\eqx
where $p(W_1=a_i,W_2=a_j)$ denotes probability of choosing in (\ref{r14})
$W_1=a_i$ and $W_2=a_j$ ($i,j=1,\ldots,m$).
Then the density moments fulfill relation 
(\ref{r3}), where intermittency exponents $\phi_q$ are equal to~:
\eq
\phi_q=\log(a_1^q p_{a_1}+\ldots+a_m^q p_{a_m}).
\lab{r17}
\eqx
The models~: $\al-$, $p-$ and $(p+\al)$ may be derived from general
multiplicative rule (\ref{r14}). To obtain the $\al-$model it is enough 
to assume random variables $W_1$, $W_2$ to be independent~:
\eqn
\langle W_1W_2 \rangle=\langle W_1 \rangle \langle W_2 \rangle.
\lab{r16}
\eqnx
The $\al-$model has no node memory therefore no conservation law can be
implemented here. 

Relation (\ref{r14}) reduces to the $p-$model after turning $m=2$ and~:
\eqn
a_2&=&1-a_1,\nonumber\\
p_{a_1}&=&p_{a_2}=0.5,\nonumber\\
p(W_2=a_2 &\mid& W_1=a_1)=1,\nonumber\\
p(W_2=a_1 &\mid& W_1=a_2)=1,
\lab{r1616}
\eqnx
where $p(W_2=a_i\mid W_1=a_j)$ denotes conditional probability of $W_2=a_i$,
if $W_1=a_j$. The $p-$model is an example of a multiplicative model with
the neighbour-node memory. Relation (\ref{r1616}) implies that the sum 
$x^{(k+1)}_{2i}+x^{(k+1)}_{2i+1}=x^{(k)}_{i}$, and the particle density 
in a node is preserved.

Finally we introduce the many parameter $(p+\al)-$model, using relation
(\ref{r14}) combined with the $\al-$ and $p-$model restrictions~:
\eqn
a_{2i}    &=&1-a_{2i-1}  ,\nonumber\\
p_{a_{2i}}&=&p_{a_{2i-1}},\nonumber\\
\eqnx
where $m$ is an even number ($i=1,\ldots,{m \over 2}$), and~:
\eqn
p(W_2=a_{2i}\,\,\,  &\mid& W_1=a_{2i-1  })  =1,\nonumber\\
p(W_2=a_{2i-1  }  &\mid& W_1=a_{2i}\,\,\,)  =1.\lab{r18}
\eqnx
One may check that the particle distribution generated in the $(p+\alpha)-$model 
is uniform. The $(p+\al)-$model may involve any number of parameters $a_i$.
Therefore it describes a more realistic case of cascading
since for large $m$ the distributions of particle density for
$W_1$ and $W_2$ (\ref{r1414}) may be approximated by a continuous
distribution function $f(x)$~: $W_{1,2}=x$ with probability $f(x)\,dx$. 
The total particle density will be preserved for any $m$, according to
(\ref{r18}).

\section{Recovering corrections in multiplicative models}

Now we calculate explicitly the correction $p_q(k)$ for 
any multiplicative model with possible neighbour-node memory. 
To do this we will split the bins ($x_i$'s) appearing in (\ref{r13}) into a left half~($i<2^{k-1}$) 
and a right half~($i\geq 2^{k-1}$) \cite{jz}~:
\eqn
p_q(k)&=&\cor{\left(\f{1}{2^k} \sum_i l_i+r_i\right)^q}=\nonumber\\
&=&\f{1}{2^{q}}\Biggl\langle \sum_{j=0}^q \left(\begin{array}{c} q\\j\end{array}\right) \left(\f{1}{2^{k-1}}\sum_i l_i\right)^j \left(\f{1}{2^{k-1}}\sum_i r_i\right)^{q-j}W_1^jW_2^{q-j}\Biggr\rangle.
\lab{r1818}
\eqnx
Using the fact that the left and right bins are independent, one arrives at
the recurrence equation:
\eq
p_q(k)=\f{1}{2^{q}}\sum_{j=0}^q \left(\begin{array}{c} q\\j\end{array}
\right) p_j(k-1) p_{q-j}(k-1) \langle W_1^jW_2^{q-j}\rangle
\lab{r19}
\eqx
which may be solved recursively together with the initial data~:
\eqn
p_q(0)&=&1,\nonumber\\
p_0(k)&=&1.
\lab{r1919}
\eqnx
A similar recurrence relation has been also obtained in \cite{pesch}.
It should be stressed that coefficients $\langle W_1^jW_2^{q-j}\rangle$
are the only parameters of the model needed to solve (\ref{r19}) 
recursively. It means that to calculate $p_q(k)$ for a given model 
we need only to know the  coefficients $\langle W_1^jW_2^{q-j}\rangle$.
In the next section we show how to apply this observation to the data
analysis.

Finally we present recovering corrections calculated for the $\al-$ and $p-$models~:
\eqn
p_q^{\al-model}(k)&=&\f{1}{2^{q}}\sum_{j=0}^q \left(\begin{array}{c} q\\j\end{array}
\right) p_j(k-1) p_{q-j}(k-1) 2^{\phi_j+\phi_{q-j}},\lab{alf}\\
p_q^{p-model}(k)  &=&2^{\phi_1\,qk} \lab{r20}
\eqnx
where $\phi_j$ denote intermittency exponent (\ref{r17}). 

\section{Implementation of the recovering corrections}

The idea how to implement the corrections $p_q(k)$ into the analysis of
the $\alpha-$model data was sketched briefly in \cite{jz}. 
Here we extend the primary scheme, and apply it to the multiplicative model
data. As it was
mentioned in the previous section, coefficients $\langle W_1^jW_2^{q-j}\rangle$ 
are the only parameters of the model needed to calculate recursively the
corrections $p_q(k)$. Let us introduce a new notation for 
$\langle W_1^jW_2^{l}\rangle$~:
\eq
\langle W_1^jW_2^{l}\rangle \equiv k_{j,l}.
\lab{not}
\eqx
We ask now how to derive $k_{j,q-j}$ from the model. One may notice
that for either $j=0$ or $l=0$ coefficients $k_{j,l}$ equal~:
\eq
k_{j,0}=k_{0,j}=2^{\phi_j},
\lab{k0}
\eqx
where $\phi_j$'s are ordinary intermittency exponents (\ref{r3})
which may be determined from relations (\ref{r44}), (\ref{r10}).
To find the value of $k_{j,l}$ ($j,l\neq 0$) we use the 
{\bf unnormalized density correlators} $c_{j,l}^{(k)}$ \cite{l1,l4,corr}~:
\eq
c^{(k)}_{j,l}=\f{1}{2^{k-1}} 
\sum_{i=0}^{2^{k-1}-1} 
\left( x^{(k)}_{2i} \right)^j \left( x^{(k)}_{2i+1} \right)^l.
\lab{defcor}
\eqx
In multiplicative models the correlators and the density moments fulfill 
the relation~:
\eq
c^{(k)}_{j,l}=z^{(k-1)}_{j+l}\cdot k_{j,l}
\lab{cz}
\eqx
which can be also rewritten as~:
\eq
\log c^{(k)}_{j,l}=(k-1)\phi_{j+l}+\log k_{j,l}.
\lab{cz2}
\eqx
Relation (\ref{cz}) may be easily derived from (\ref{r14})~: since 
each term in sum (\ref{defcor}) originates from one node $x^{(k-1)}_i$, 
it can be rewritten as~:
{\footnotesize $\left( x^{(k)}_{2i} \right)^j\left(x^{(k)}_{2i+1}\right)^l=\left(x^{(k-1)}_i\right)^{j+l} W_1^j W_2^l$}.
Relations (\ref{cz}), (\ref{cz2}) imply that we may derive $k_{j,l}$
in a straigthforward way by calculating correlators and density moments 
from data, and applying to them the standard $\chi^2-$fit.

Applying the standard method to the correlators at the
previous cascade stages, we expect to find the discrepancy (due to the 
recovering procedure) between reconstructed correlators and the true ones,
similarly as for the density moments.
It can be proved (see Appendix B) that the discrepancy may be expressed 
in terms of recovering correction $p_q(k)$ (\ref{r13})~:
\eq
c^{(n-k)}_{j,l;\,rec.}=c^{(n-k)}_{j,l}p_{j+l}(k).
\lab{crec}
\eqx  
Now we have all tools needed for implementation of recovering corrections
into the multiplicative data analysis. Below we propose an implementation 
algorithm which recursively adjusts the primary parameters 
$\phi_q$, $k_{j,l}$ ($j+l=q$, $jl>0$) obtained after applying the standard 
method to the data~:\\

({\bf INPUT}) parameters $\tdm\phi_1,\ldots,\tdm\phi_{q-1}$,
$\tdm k_{j,l}$ ($j+l=1,\ldots,q-1$) 
obtained after applying the implementation algorithm for 
$q=1,2,\ldots,q-1$ step-by-step\\
(for determination of $\phi_1$ see Appendix C.3),\\

({\bf 1}) derive $\tdm\phi'_q$, $\tdm k'_{j,q-j}$ ($j=1,\ldots,q-1$) 
from data, using the standard method\\ i.\ e.\ reconstruct the cascade 
using (\ref{r5}) and derive the parameters from relations~:
\eq
\label{al1}
\log z^{(k)}_{q;\,rec.\ } = k\cdot \phi'_q + b.
\eqx
\eq
c^{(k)}_{j,l;\,rec.\ }=z^{(k-1)}_{j+l;\,rec.\ }\cdot k'_{j,l}
\lab{al2}
\eqx
where $k=1,\ldots,n$ (cf.\ (\ref{r44}), (\ref{cz})),\\

({\bf 2}) derive $\tdm\phi_{q;\,corr.\ }$, $\tdm k_{j,q-j;\,corr.\ }$
($j=1,\ldots,q-1$) in the following substeps~:\\

\noindent
({\small \bf 2.0}) calculate $p_q(k)$ from relation (cf.\ (\ref{r19}))~:
\eqn
p_q(k)=\f{1}{2^{q}}\sum_{j=0}^q \left(\begin{array}{c} q\\j\end{array}
\right) p_j(k-1) p_{q-j}(k-1) k_{j,q-j},
\lab{al3}
\eqnx
using $\tdm\phi'_q$, $\tdm k'_{j,q-j}$ derived in step
({\bf 1}), and estimate $\tdm\phi_{q;\,corr.}$ from (cf.\ (\ref{r10}))~:
\eqn
\label{al4}
\log z^{(n-k)}_{q;\,rec.} -\log(p_q(k)) = (n-k)\cdot \phi_{q;\,corr.} +b,
\eqnx\\

\noindent
({\small \bf 2.1} ) calculate $p_q(k)$ from (\ref{al3}) using $\phi_{q;\,corr.}$ (other parameters
as after step (1)), and estimate $\tdm k_{1,q-1;\,corr.}$ from relation
(cf.\ (\ref{cz2}),(\ref{crec}), see also Appendix C.4)~:
\eqn
\log c^{(n-k)}_{j,l;\,rec.}-\log(p_{j+l}(k))=(n-k-1)\phi_{j+l}+\log k_{j,l;\,corr.},
\lab{al5}
\eqnx

\noindent
$\ldots$,\\

\noindent
({\small \bf 2.q-1}) calculate $p_q(k)$ from (\ref{al3}), using all previously derived parameters
$\tdm\phi_{q;\,corr.\ }$, $\tdm k_{j,q-j;\,corr.\ }$, and 
estimate $\tdm k_{q-1,1;\,corr.}$ from (\ref{al5}),\\

({\bf 3}) compare the values of $\tdm\phi'_q$, $\tdm k'_{j,q-j}$ and 
    $\tdm\phi_{q;\,corr.}$, $\tdm k_{j,q-j;\,corr.}$ ($j=1,\ldots,q-1$). If the relative
    difference is large, assume~:
\eqn
\phi'_q   &:=&\phi_{q;\,corr.},\nonumber\\ 
k'_{j,q-j}&:=&k_{j,q-j;\,corr.}\nonumber
\eqnx
and repeat steps (2),(3) recursively until the relative difference
between parameters before and after step (2) is small enough. Then
go to the output, assuming $\tdm\phi_q:=\tdm\phi'_q$, 
$\tdm k_{j,q-j}:=\tdm k'_{j,q-j}$   \\

({\bf OUTPUT}) parameters $\tdm\phi_1,\ldots,\tdm\phi_{q}$; 
$\tdm k_{j,l}$ ($j+l=1,\ldots,q$).\\

Other techniqual details and problems which may appear when applying 
the algorithm to data are listed in Appendix C.\\

We have performed numerical simulations of the $\alpha-$, $p-$ and
$(p+\al)-$models in order to test how the implementation 
algorithm works in practice.
We generated 10000 cascades of the 10 step length for the $\alpha-$ 
and $(p+\alpha)-$models, and one cascade
of the 10 step length for the $p-$model \footnote{it can be proved that for a given
parameter set the $p-$model generates always the same values of the correlators
and density moments} for two different parameter sets separately.

Implementation algorithm analized the data of the last cascade step.
For each event it estimated the value of normalized intermittency exponents
$\phi_{2;\,norm.}$, $\phi_{3;\,norm.}$ ($\phi_{i;\,norm.}:=\phi_i-i\cdot\phi_1$),
using the standard  method (step 1) with recovering corrections included 
(steps 2,3).
The results are presented in Figs.\ 1, 2, 3, 4 
(for the $\al-$ and $(p+\al)-$models)
and in Tabs.\ 1, 2.

For the $\al-$model the histograms of $\phi_{2;\,norm.}$, $\phi_{3;\,norm.}$ 
obtained in the standard method and the histograms with recovering corrections
included are almost identical. In this case the recovering corrections can
be implemented better when one applies directly dedicated $\alpha-$model recovering 
correction $p_q^{\al-model}(k)$ (\ref{alf}) (see Figs.\  1, 2 and Tab.\ 1).   
The different accuracy of these both approaches is due to the fact that 
the random variables $W_1$, $W_2$ are independent in the $\alpha-$model, 
and the coefficients $k_{j,q-j}$ (\ref{not}) are approximated better by 
the product $2^{\phi_j}\cdot 2^{\phi_{q-j}}$ than from correlators (\ref{cz}).  

On the contrary, the implementation algorithm works well for  
the $(p+\alpha)-$model (see Figs.\ 3, 4 and Tab.\ 2). 
For the $(p+\alpha)-$model the histogram with the recovering corrections 
included approximates well the theoretical value of normalized 
intermittency exponent. The histogram obtained by using the standard 
method is moved slightly to the left in comparison to the histogram with
recovering corrections included. 
 
We have checked that for the $p-$model the theoretical values of normalized 
intermittency exponents are estimated perfectly by both standard method 
and implementation algorithm, as we have expected 
\footnote{it follows from relations (\ref{r8}),(\ref{r20}) that in the $p-$model 
$\phi_{i;\,rec.\,,norm.}=\phi_{i;\,norm.}$ if $i>1$}. 

It should be also mentioned that the histograms generated by both
implementation algorithm and dedicated recovering corrections
are symmetric, in contrast to the standard ones, and their dispersions are relatively
small (see Tabs.\ 1, 2). 
Finally, we notice that the accuracy of the estimation of intermittency 
exponents, while applying the standard method and the improved one to a given 
model, depends on the parameters of this  model (cf. Figs.\ 1, 2 and Figs.\ 3, 4).
In Appendix D we present a qualitative analysis of the effect for intermittency
exponents of the second rank. A similar analysis has also been done
in \cite{pesch}. 

\section{Conclusions}

To sum up we analized the estimation of intermittency exponents from the data
which were generated by a multiplicative random cascading process.
The following methods were applied: the standard method of cascade
recovering (\ref{r5}) and the improved method which included
recursively the recovering corrections. The improved method was applied
in the form of the implementation algorithm. Numerical simulations have been
performed to check how both methods work in practice.
The conclusions may be summarized as follows~:\\

(a)  
standard method of estimation of intermittency exponents does not apply
for the whole class of multiplicative models~: its accuracy
depends on the specific properties of the model 
and its parameters. The method does not detect a conservation law
if present in the model;\\ 

(b)
improved method of estimation of intermittency exponents applied in the
form of recursive implementation algorithm either corrects the standard
method estimation or does not change the standard method result.
In the latter case the estimation 
may be corrected by applying the dedicated recovering corrections.
In any case the improved method tests the applicability of the standard method, and allows
one to estimate the accuracy of the intermittency exponents estimation.

While applying the improved method, the parameters of the model are 
adjusted recursively from the primary (standard method) parameters.
The histograms generated by the improved method are symmetric, 
and their dispersions are of the same order as those determined for 
the standard method. The improved method takes into account the
neighbour-node memory (a conservation law) if present in the model,
by including the density correlators into the estimation scheme.
\\   


\section*{Acknowledgements}

I would like to thank Prof.\ A. Bia{\l}as for reading the manuscript and
many suggestions and comments and Dr.\ R.\ Janik for discussions. 
This work was supported in part by Polish Government grant 
Project (KBN) 2P03B04214.
\section*{Appendix A}

We prove relation (\ref{r13}). 
The density moment $z^{(n-k)}_{q;\,rec.}$
may be rewritten as~:
\eq
z^{(n-k)}_{q;\,rec.}=\f{1}{2^{n-k}} \sum_{i=0}^{2^{n-k}-1}  \left(
y^{(n-k)}_i \right)^q=
\f{1}{2^{n-k}} \sum_{i=0}^{2^{n-k}-1}  
\left(\f{1}{2^k} \sum_{j=0}^{2^k-1}x^{(n)}_{2^k\times i+j} \right)^q.
\lab{ap0}
\eqx
Notice that~:
\eq
x^{(n)}_{2^k\times i+j}=x^{(n-k)}_i\cdot x^{(k)}_j.
\lab{ap1}
\eqx
Substituting (\ref{ap1}) into (\ref{ap0}), one arrives at~:
\eq
z^{(n-k)}_{q;\,rec.}=
\f{1}{2^{n-k}} \sum_{i=0}^{2^{n-k}-1} \left( x^{(n-k)}_i \right)^q   
\left(\f{1}{2^k} \sum_{j=0}^{2^k-1}x^{(k)}_{j} \right)^q=
z^{(n-k)}_q\,p_q(k).
\eqx
%

\section*{Appendix B}

We prove relation (\ref{crec}). The correlator $c^{(n-k)}_{j,l;\,rec.}$
may be rewritten as~:
{\footnotesize
\eqn
c^{(n-k)}_{j,l;\,rec.}&=&\f{1}{2^{n-k-1}} \sum_{i=0}^{2^{n-k-1}-1}  
\left(y^{(n-k)}_{2i} \right)^j \left(y^{(n-k)}_{2i+1} \right)^l\nonumber\\
&=&\f{1}{2^{n-k-1}} \sum_{i=0}^{2^{n-k-1}-1}  
\left(\f{1}{2^k} \sum_{m=0}^{2^k-1}x^{(n)}_{2^k\times 2i+m} \right)^j
\left(\f{1}{2^k} \sum_{r=0}^{2^k-1}x^{(n)}_{2^k\times (2i+1)+r} \right)^l.
\lab{app0}
\eqnx
}
Relation (\ref{ap1}) implies~:
\eq
c^{(n-k)}_{q;\,rec.}=
\f{1}{2^{n-k-1}} \sum_{i=0}^{2^{n-k-1}-1} 
\left( x^{(n-k)}_{2i} \right)^j
\left( x^{(n-k)}_{2i+1} \right)^l
\left(\f{1}{2^k} \sum_{j=0}^{2^k-1}x^{(k)}_{j} \right)^{j+l}=
c^{(n-k)}_{j,l}\,p_{j+l}(k).
\eqx
%

\section*{Appendix C}

We list some techniqual details which can be useful when applying the
implementation algorithm to data, and discuss possible problems.\\

({\bf 1})
The recovering corrections applied for calculating
$\phi_2$, $\phi_3$ read~:
{\footnotesize
\eqn
p_1(k)&=&2^{k\phi_1}\lab{p1}\\
p_2(k)&=&{1 \over 4}(\,p_2(k-1)k_{2,0}+p_2(k-1)k_{0,2}+2p_1^2(k-1)k_{1,1}\,)
\lab{p2}\\
p_3(k)&=&{1 \over 8}(\,p_3(k-1)k_{3,0}+p_3(k-1)k_{0,3}\nonumber\\
      &+& 3p_1(k-1)p_2(k-1)k_{1,2}+3p_2(k-1)p_1(k-1)k_{2,1}\,)\lab{p3}
\eqnx
}\\

({\bf 2})
Since random variables $W_1$, $W_2$ generate the uniform 
distribution (cf. (\ref{unif})) relations (\ref{p1},\ref{p2},\ref{p3}) may
be simplified by substituting~:
\eq
k_{j,l}=k_{l,j}.
\eqx
The (experimental) estimation of $k_{j,l}$ will be better
if we determine $c_{j,l}^{(n)}$ as~:
\eq
c^{(n)}_{j,l}=\f{1}{2^{n-1}} 
\sum_{i=0}^{2^{n-1}-1}
\left\{ 
\left( x^{(n)}_{2i} \right)^j \left( x^{(n)}_{2i+1} \right)^l+
\left( x^{(n)}_{2i} \right)^l \left( x^{(n)}_{2i+1} \right)^j
\right\}
\lab{cor}.
\eqx\\

({\bf 3})
Determination of $\phi_1$. It follows from relations (\ref{r8}),(\ref{p1})
that the unnormalized reconstructed density moment $z_{1;\,rec.\ }$
for any multiplicative model takes the form~:
\eq
z_{1;\,rec.\ }^{(k)}=2^{n\phi_1}=z_1^{(n)},
\lab{z1}
\eqx
which does not depend on the cascade stage $k$. 
Therefore the proposed estimation of $\phi_1$, implied by (\ref{z1})
reads~:
\eq
\phi_1={\log z_{1;\,rec.\ }^{(n)} \over n}.
\lab{phi1}
\eqx

({\bf 4})
Estimation of coefficients $k_{j,l}$ in the implementation algorithm.
To estimate coefficients $k_{j,l}$ in the step (1) of the implementation
algorithm we calculate the reconstructed  correlators and density
moments, and apply relation (\ref{al2}) whereas in the step (2) 
we use relation (\ref{al5}) to do the same. 

The brief explanation of the method is following. 
In formula (\ref{al2}) $k_{j,l}$ appears as a slope, and an ordinary 
linear $\chi^2$-fit may estimate it with a good accuracy. 
This approach works well for the reconstructed correlators and moments.

On the contrary,  the long-range correlations: $\log b$ present in 
relation (\ref{al5}) add to the value of $\log k_{j,l}$,
and generate a large error while estimating $k_{j,l}:=exp(\log k_{j,l}+\log b)$
from (\ref{al5}). We could try to estimate $b$, assuming that~:
\eq
z^{(n)}_q=2^{n\cdot \phi_q}b.
\lab{ap4}
\eqx
Then it follows from (\ref{al5}) that~:
\eq
\log c^{(n)}_{j,l}=(n-1)\phi_{j+l}+\log {\bar k_{j,l}},
\lab{ap5}
\eqx
and $\log k_{j,l}$ equals~:
\eq
\log k_{j,l}=\log {\bar k_{j,l}} - \log b.
\lab{ap6}
\eqx
The latter approach does not work for the reconstructed moments, 
where relations (\ref{ap4},\ref{ap5}) apply only approximately. 
However, it applies 
quite well for the moments with recovering corrections included because 
for this case relation (\ref{al2}) would require including recovering corrections 
to both correlators and density moments which in turn would generate 
larger error in estimating $k_{j,l}$. 

It was checked that the above method works for the multiplicative model
data. However, the problem of determination of coefficients $k_{j,l}$ 
and, in particular, the problem of the determination of correlators 
$c^{(n)}_{j,l}$ from the real data is much more complicated 
(see e.\ g.\ \cite{l4, corr}). In this case the method needs some
improvement which we will not discuss here.\\  

({\bf 5})
The recursive implementation algorithm is not always convergent. 
Since estimation at the $q$th step depends upon parameters which were 
adjusted in the previous steps $1,\ldots,q-1$, 
the estimation errors propagate and get larger with growing $q$. 
Then it happens sometimes (not very often), that recursive adjusting ends 
with the repeating a sequence of different values of 
parameters, or parameters become not  definite. In such a case we stop 
the algorithm, assuming for the values of intermittency parameters 
those derived in step (1).
\section*{Appendix D}

The accuracy of the estimation of intermittency exponents, while applying 
the standard method and the improved one to a given model, 
depends on the parameters of this  model (cf. Figs.\ 1, 2 and Figs.\ 3, 4).
Below we present a qualitative analysis of the effect for the intermittency 
exponents of the second rank.

Equation (\ref{p2}) may be solved analytically, and the solution 
(valid for any multiplicative model) takes the closed form~:
\eq
p_2(k)=(1-A)\cdot 2^{(\phi_2-1)k}+A\cdot 2^{2k\phi_1},
\lab{rp2}
\eqx
where~:
\eq
A={ k_{1,1} \over 2^{2\phi_1+1}-2^{\phi_2} }.
\lab{rp3}
\eqx
The reconstructed moments then read~:
\eqn
z_{2;\,rec.\ }^{(k)}&=&(1-A)\,
2^{(\phi_2-1) n}\cdot 2^k + A\,2^{\phi_1n}\,\cdot 2^{\phi_{2;\,norm.\ } k}.
\lab{rz2}
\eqnx
There are {\bf two} power law terms~: $2^k$ and $2^{\phi_{2;\,norm.\ }k}$ 
in  $z_{2;\,rec.\ }^{(k)}$. In order to establish how they influence 
the determination of $\phi'_{2}$ (\ref{r44},\ref{al1}) we performed 
the following check. For a given multiplicative model with fixed parameters
(e.\ g.\ $\alpha-$ model with parameters as in Figs.\, 1, 2) 
we established  the values of $z_{2;\,rec.\ }^{(k)}$  from (\ref{rz2}), 
and made the linear $\chi^2-$fit to the points $(k, \log z_{2;\,rec.\ }^{(k)})$.
The slope obtained from the fit estimated the value of $\phi'_{2}$. 
For the $\alpha-$model we have obtained~: for case a) 
$\phi'_2=0.0250$, and for case b) $\phi'_2=0.284$. Both results agree with the
$\phi'_2$'s obtained from the model simulation (cf. Figs. 1,2 and Tab.\ 1).
Similar analysis can be also peformed for the $(p+\alpha)-$model.

The above results confirm our observation that the accuracy of the estimation 
of intermittency exponents from the standard method depends on the
parameters of the model. 
And, if the estimation of primary parameters $\phi'_q$, $k'_{j,l}$ 
is more accurate, their recursive adjusting performed by the implementation 
algorithm will be faster and more accurate as well. It means that in this
case the improved analysis works  better as well. 

\newpage
\noindent
{\bf Figure captions}\\ \\
\noindent
{\bf Figs.\ 1, 2}
Estimation of normalized intermittency exponents $\phi_{2;\,norm.}$ 
and $\phi_{3;\,norm.}$ 
for $\al-$model, using the standard  method (dotted line), the improved method 
with the implementation algorithm (thin solid line), and dedicated 
$\al$-~corrections (\ref{alf}) (dashed line) compared with the theoretical
values (solid line), performed for two sets of $\al-$model parameters~:\\ \\
a) $a_1=0.8$, $a_2=1.1$, $p_1=1/3$\\
b) $a_1=0.5$, $a_2=1.5$, $p_1=1/2$.\\ \\
{\bf Figs.\ 3, 4}
Estimation of normalized intermittency exponents $\phi_{2;\,norm.}$ 
and $\phi_{3;\,norm.}$ 
for $(p+\al)-$model, using the standard  method (dotted line), 
the improved method with the implementation algorithm (thin solid line), 
compared with the theoretical values (solid line), performed for
two sets of $(p+\al)-$model parameters~:\\ \\
a)    $a_{2i}=1-a_{2i-1}$, $p_{2i}=p_{2i-1}=0.05$ for $i=1,\ldots,10$,\\
      $a_1   =0.2$, $a_3   =0.5$, $a_5   =0.6$, $a_7   =0.3$, $a_9   =0.45$,\\
      $a_{11}=0.25$, $a_{13}=0.1 $, $a_{15}=0.15$, $a_{17}=0.87$, $a_{19}=0.66$,\\ \\
b)    $a_{2i}=1-a_{2i-1}$, $p_{2i}=p_{2i-1}$ for $i=1,\ldots,10$,\\
      $a_1   =0.2$, $a_3   =0.5$, $a_5   =0.6$, $a_7   =0.3$, $a_9   =0.45$,\\
      $a_{11}=0.25$, $a_{13}=0.1 $, $a_{15}=0.15$, $a_{17}=0.87$, $a_{19}=0.66$,\\ \\
      $2 p_1=0.05$, $2 p_3=0.15$, $2 p_5=0.25$, $2 p_7=0.40$, $2 p_9=0.05$,\\
      $2 p_{11}=0.05$, $2 p_{13}=0.02$, $2 p_{15}=0.02$, $2 p_{17}=0.005$, $2 p_{19}=0.005$,\\ \\
\newpage
\begin{table}[hbpt]
\noindent
{\bf Tab.\ 1} Estimation of normalized intermittency exponents 
$\phi_{2;\,norm.}$ and $\phi_{3;\,norm.}$ and their dispersions 
for the $\alpha-$model, using the standard  method (second column), 
the improved method with the implementation algorithm (third column), 
and dedicated  $\alpha-$~corrections (\ref{alf}) (fourth column), 
compared with the theoretical values (first column), performed for 
two sets of $\alpha-$model parameters (see Figs.\ 1, 2)~:

\begin{center}
\begin{tabular}{|r|c|c|c|c|c|c|}
\hline \hline
                & theor.&standard&algorithm& $\alpha$-corr.\\
\hline
a) $\phi_{2;\,norm.}$   &$0.0285$&$0.0251\pm0.004$&$0.0246\pm0.0033$&$0.0288\pm0.004$\\
\hline
   $\phi_{3;\,norm.}$   &$0.0813$&$0.0757\pm0.010$&$0.0727\pm0.009$&$0.0798\pm0.0111$\\
\hline
\hline
b) $\phi_{2;\,norm.}$   &$0.322 $&$0.264\pm0.044$ &$0.253\pm0.050$&$0.276\pm0.051$\\
\hline
   $\phi_{3;\,norm.}$   &$0.807 $&$0.653\pm0.105$ &$0.666\pm0.125$&$0.750\pm0.131$\\
\hline \hline
\end{tabular}
\end{center}
\end{table}
\begin{table}[hbpt]
\noindent
{\bf Tab.\ 2} Estimation of normalized intermittency exponents 
$\phi_{2;\,norm.}$ and $\phi_{3;\,norm.}$ and their dispersions 
for the $(p+\alpha)-$model, using the standard  method (second column), 
the improved method with the implementation algorithm (third column), 
compared with the theoretical values (first column), performed for 
two sets of $(p+\alpha)-$model parameters (see Figs.\ 3, 4)~: 

\begin{center}
\begin{tabular}{|r|c|c|c|c|c|c|}
\hline \hline
                & theor.&standard&algorithm\\
\hline
a) $\phi_{2;\,norm.}$   &$0.333$&$0.322\pm0.044$ &$0.305\pm0.066$\\
\hline
   $\phi_{3;\,norm.}$   &$0.832$&$0.736\pm0.118$ &$0.720\pm0.174$\\
\hline
\hline
b) $\phi_{2;\,norm.}$   &$0.177$&$0.170\pm0.023$ &$0.173\pm0.029$\\
\hline
   $\phi_{3;\,norm.}$   &$0.478$&$0.438\pm0.069$ &$0.470\pm0.092$\\
\hline \hline
\end{tabular}
\end{center}
\end{table}
%
%
\noindent
\begin{figure}[t]
\epsfig{width=15cm, file=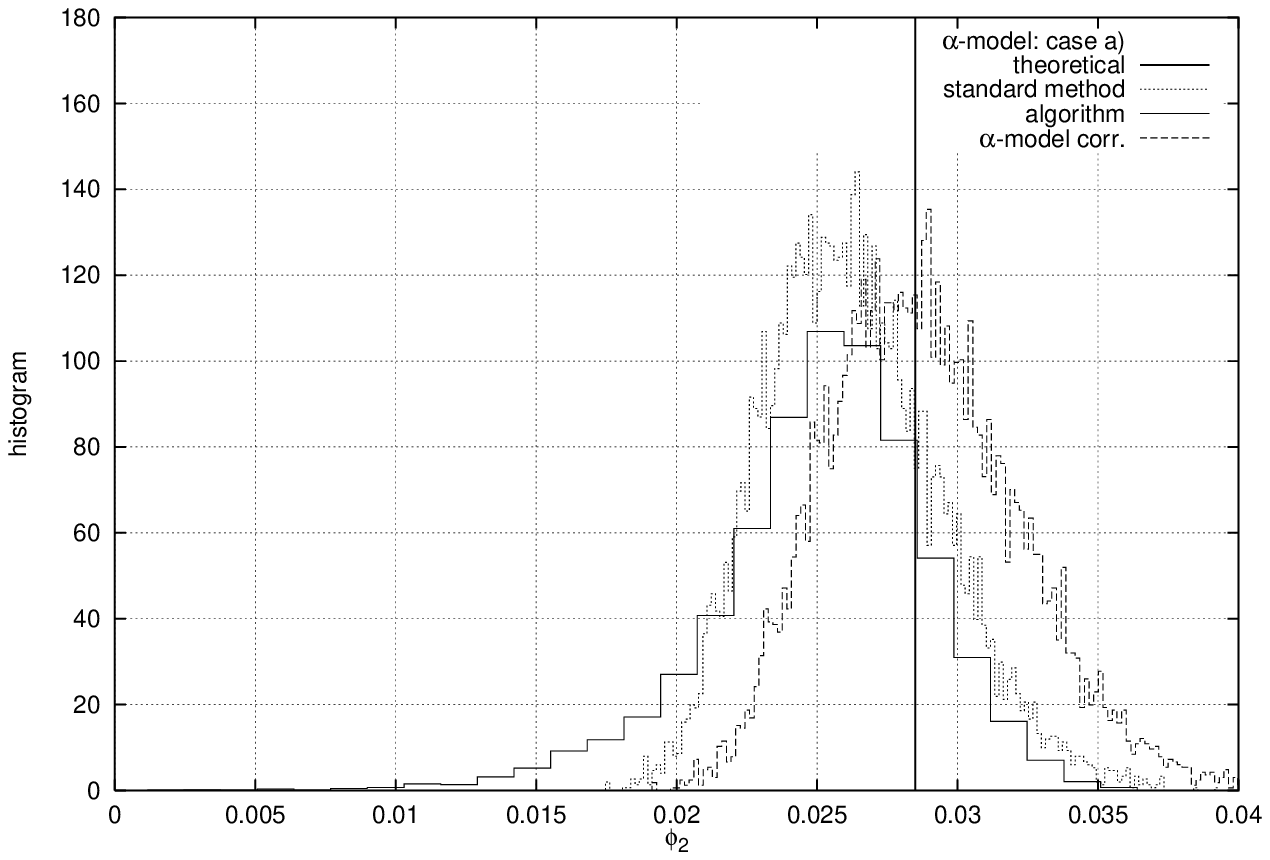}\hfill\mbox{}\\
\epsfig{width=15cm, file=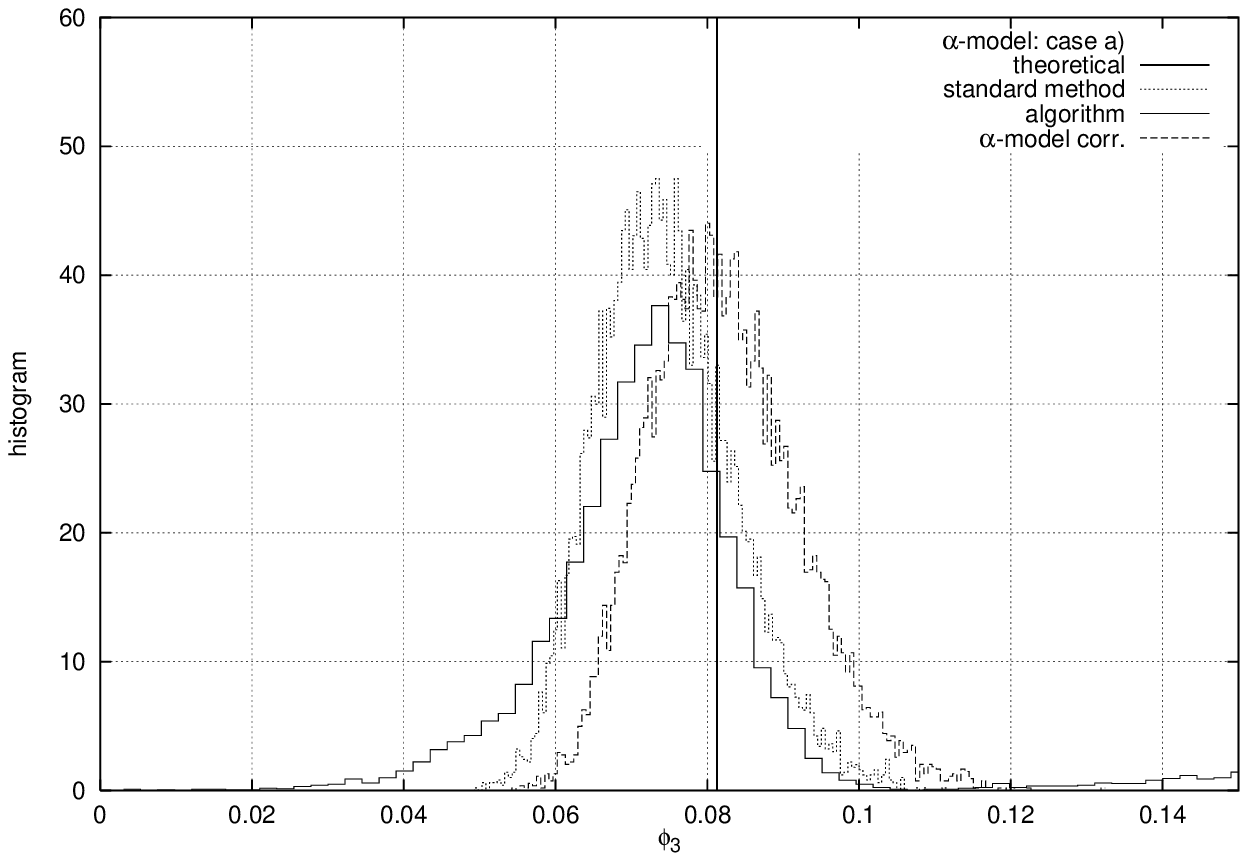}\hfill\mbox{}\\
\caption{}
\label{fig1}
\end{figure}
\noindent
\begin{figure}[t]
\epsfig{width=15cm, file=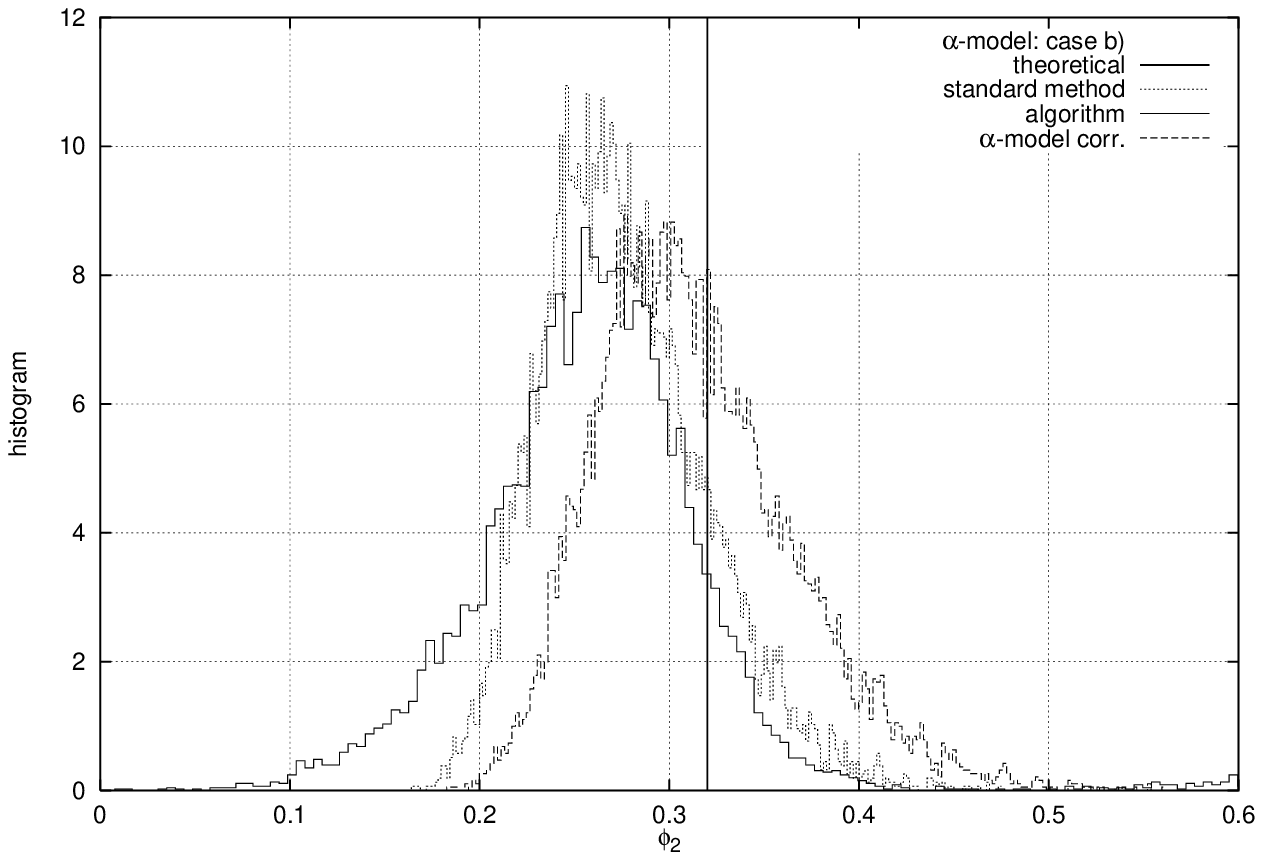}\hfill\mbox{}\\
\epsfig{width=15cm, file=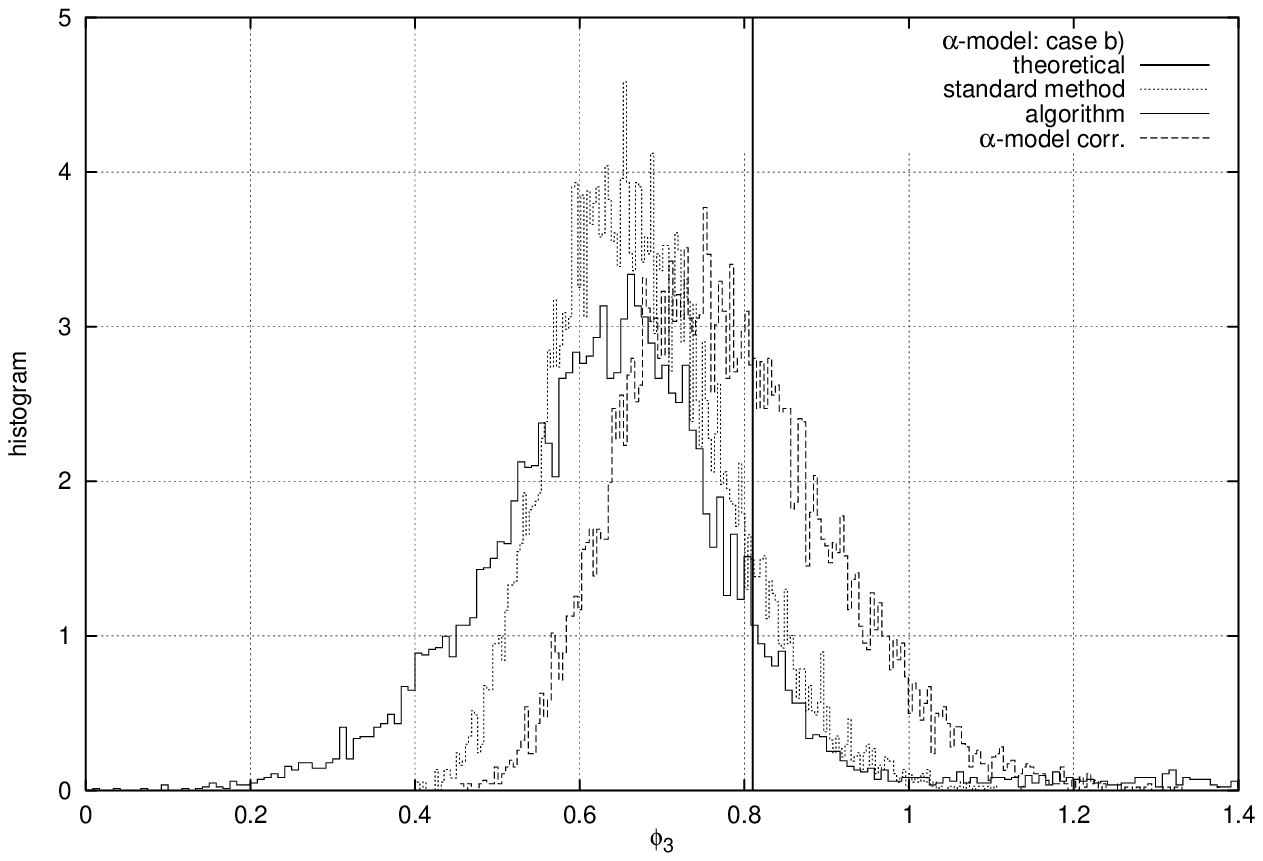}\hfill\mbox{}\\
\caption{}
\label{fig2}
\end{figure}
\noindent
\begin{figure}[t]
\epsfig{width=15cm, file=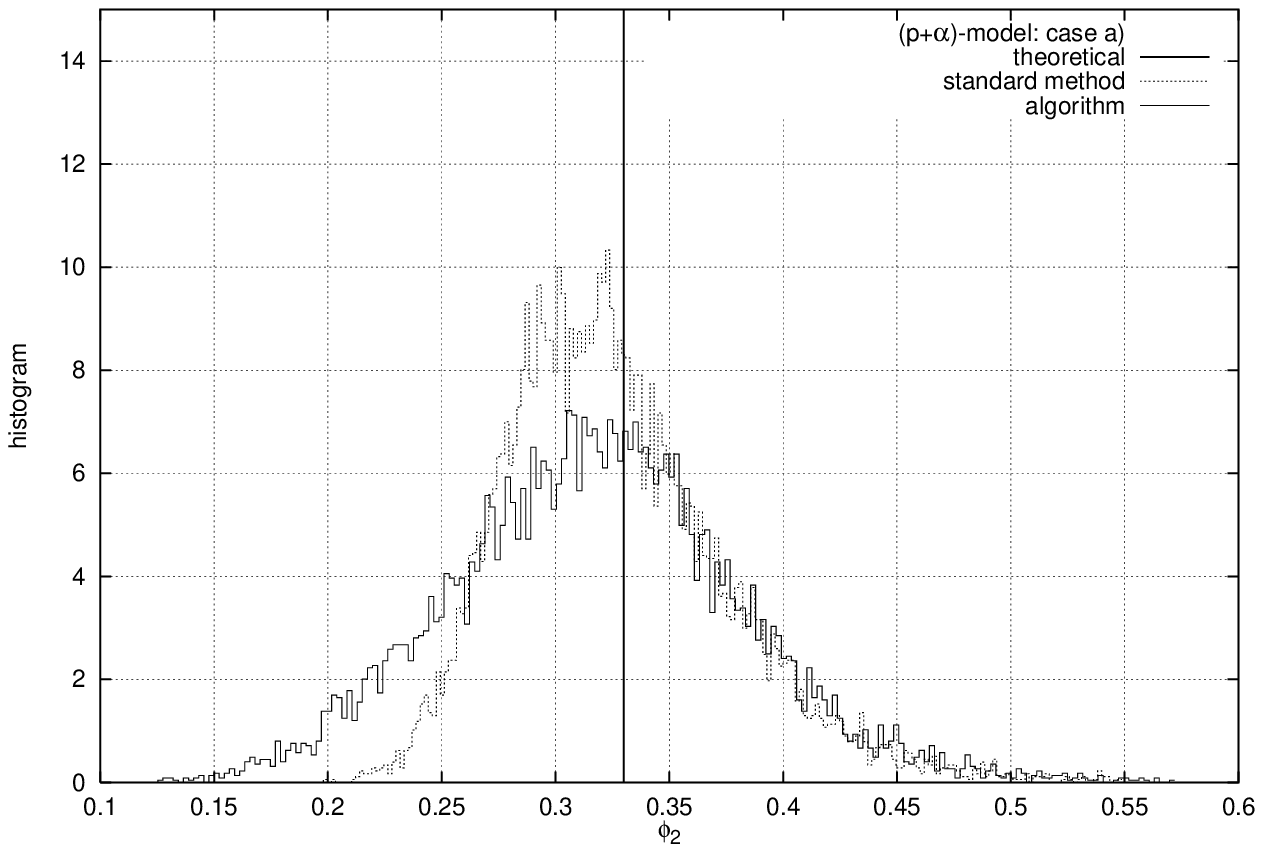}\hfill\mbox{}\\
\epsfig{width=15cm, file=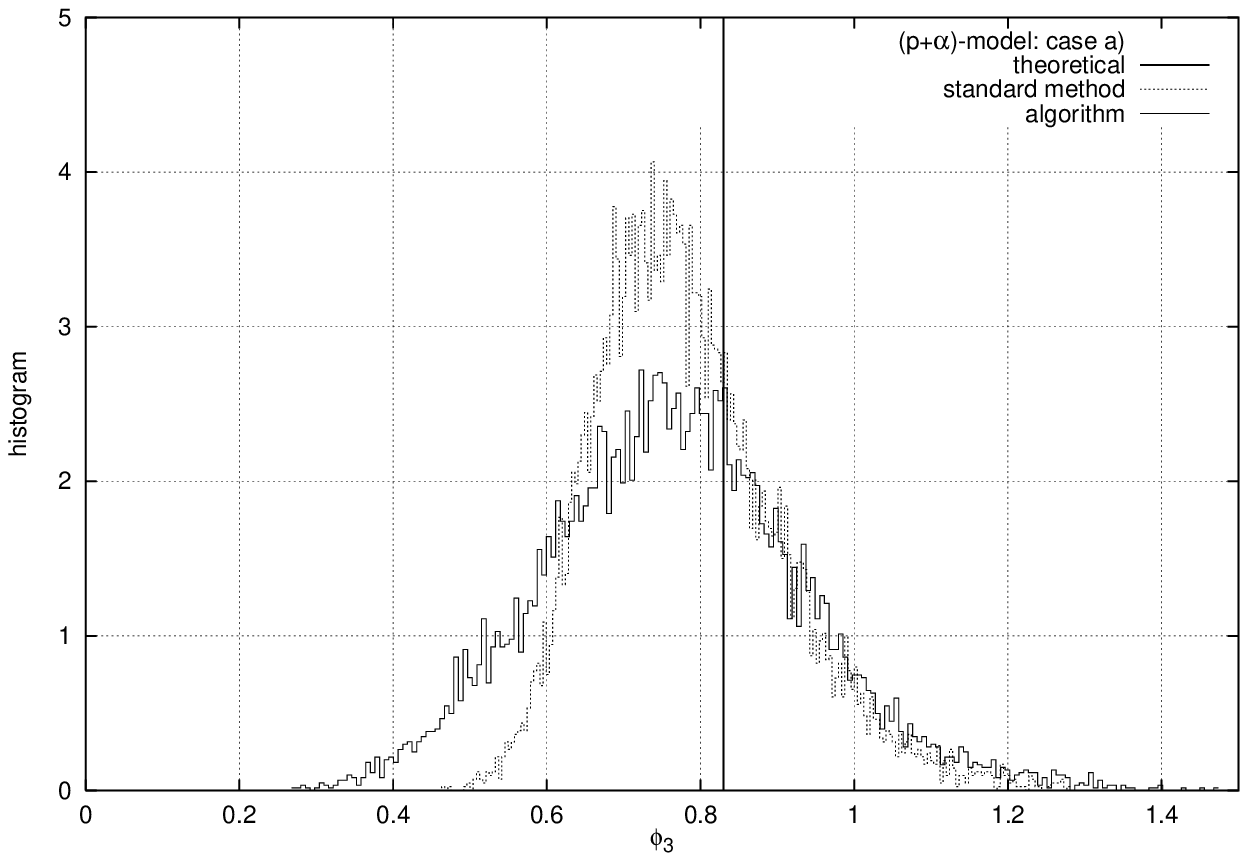}\hfill\mbox{}\\
\caption{}
\label{fig3}
\end{figure}
\noindent
\begin{figure}[t]
\epsfig{width=15cm, file=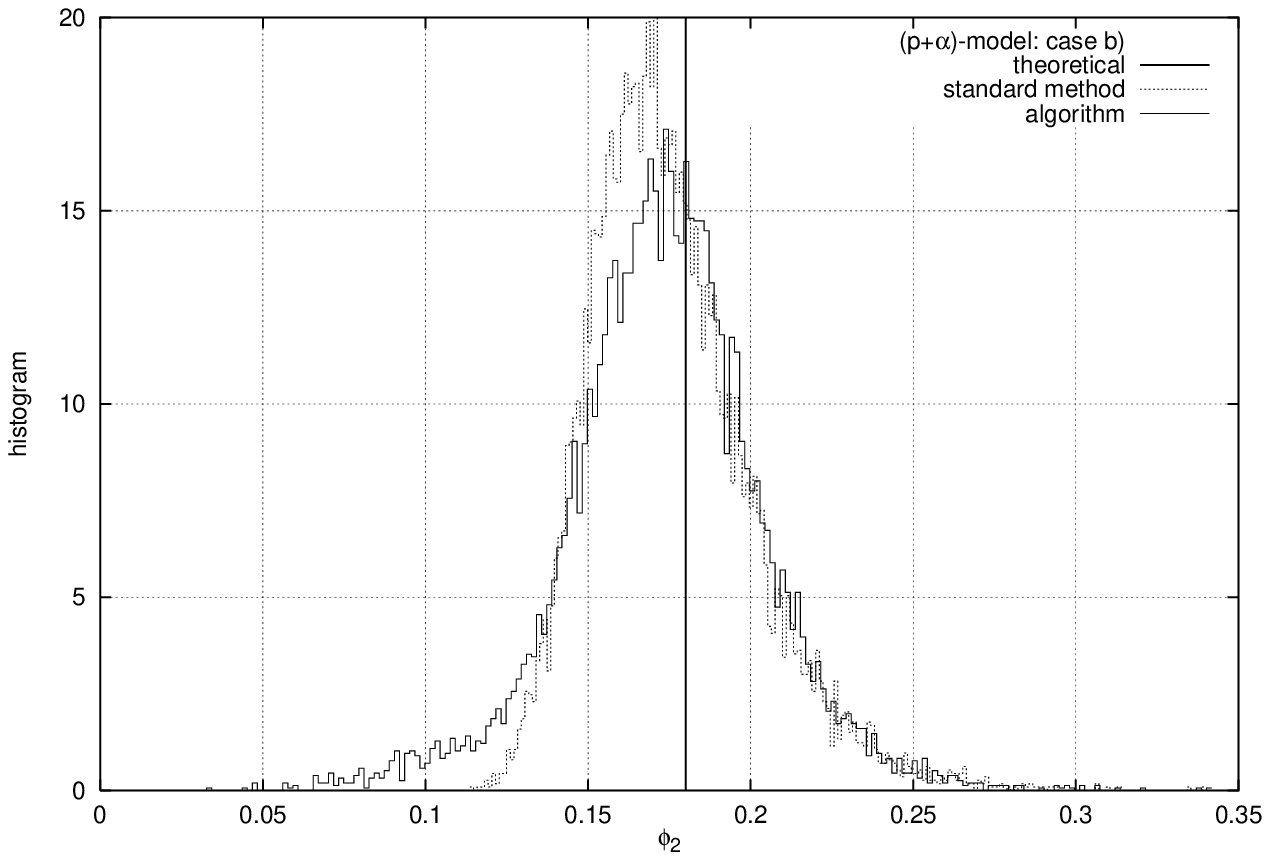}\hfill\mbox{}\\
\epsfig{width=15cm, file=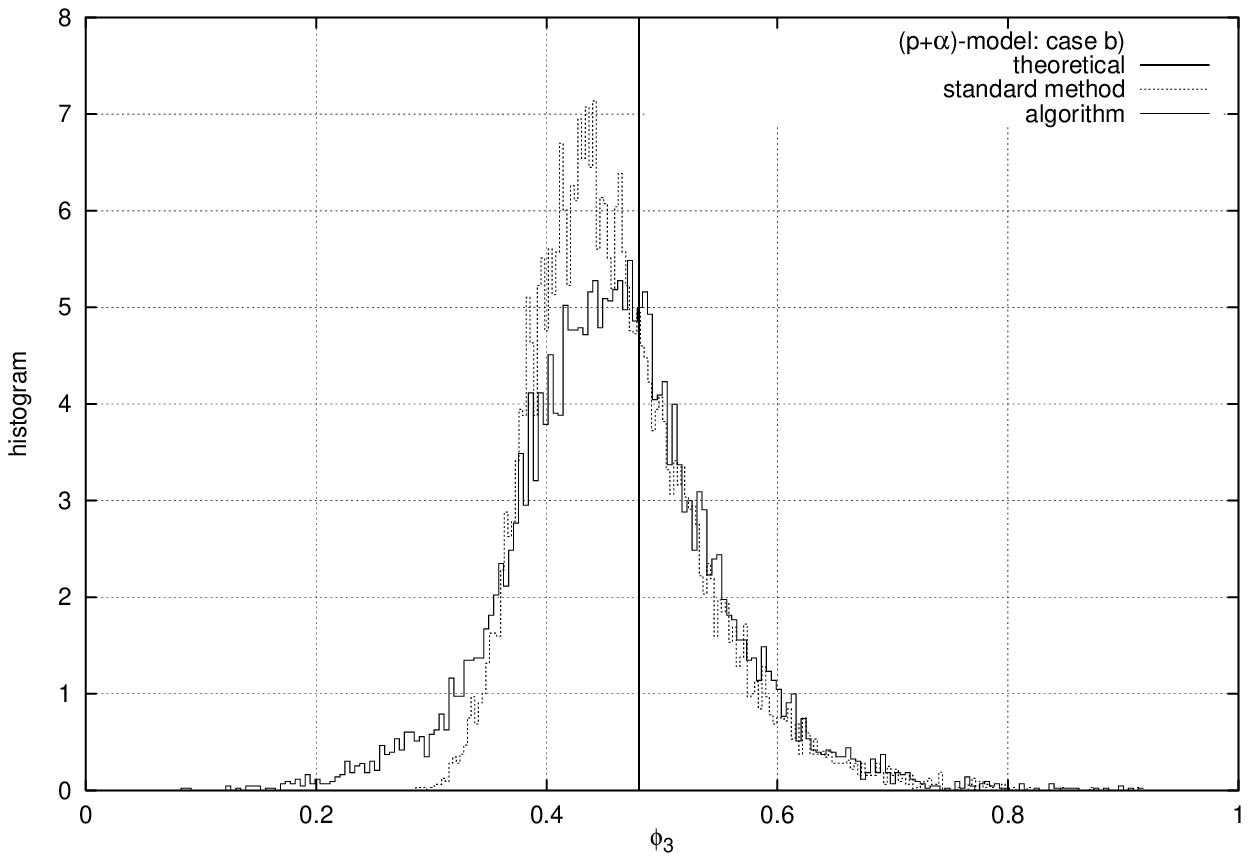}\hfill\mbox{}\\
\caption{}
\label{fig4}
\end{figure}
%
%
%

\begin{thebibliography}{3}


\bibitem{l1} A.\ Bialas, R.\ Peschanski, {\sl Nucl.\ Phys.\ }{\bf B273}, 703
(1986); A.\ Bialas, R.\ Peschanski, {\sl Nucl.\ Phys.\ }{\bf B207}, 59
(1988)
%
\bibitem{l2} JACEE coll.\, T.\ H.\ Burnett et al.\ , {\sl Phys.\ Rev.\
Lett. }{\bf 50} (1983) 2062,
%
\bibitem{acc} P.\ Carlson (UA5), 4th Topical Workshop on P-P Collider
Physics, Bern, March 1983;
G.\ J.\ Alner et al., {\sl Phys.\ Rep.\ }{\bf 154}, (1987) 247,
%
\bibitem{l4} for recent reviews see \\
             E.\ A.\ de Wolf, I.\ M.\ Dremin, W.\ Kittel, \, {\sl
Phys.\ Rep.\ 270 }(1996) 1,\\
             P.\ Bozek, M.\ Ploszajczak, R.\ Botet, {\sl Phys.\
Rep.\ }{\bf 252} (1995) 101,
%
\bibitem{chin} A.\ Bialas, R.\ Hwa, {\sl Phys.\ Lett.\ }{\bf B253} (1991) 436;
L.\ Lianshou, F.\ Jinghua, W.\ Yuanfang, {\sl HZPP} {\bf
9807},
%
\bibitem{models} 
R.\ Peschanski, {\sl Nucl.\ Phys.\ }{\bf B327} (1989) 144;
P.\ Brax, R.\ Peschanski, {\sl Nucl.\ Phys.\ }{\bf B346} (1990) 65;
A.\ Bialas, R.\ Peschanski, {\sl Phys.\ Lett.\ }{\bf B207} (1988) 59;
H.\ C.\ Eggers, M.\ Greiner, P.\ Lipa, {\sl Phys.\ Rev.\ Lett.\ }{\bf 80}
(1998) 5333; 
H.\ C.\ Eggers, M.\ Greiner, P.\ Lipa, {\sl hep-ph 9811204},
%
\bibitem{bsz} A.\ Bialas, A.\ Szczerba, K.\ Zalewski, {\sl Z.\
Phys. }{\bf C46} (1990) 163,
%
\bibitem{pesch} P.\ Desvallees, R.\ Ouziel, R.\ -Peschanski,
{\sl Phys.\ Lett.\ }{\bf B235} (1990) 317; Y.\ Gabellini, 
J.\ -L.\ Meunier, R.\ Peschanski, {\sl Z.\ Phys.\ }{\bf C55} (1992), 455,
%
\bibitem{hwa} R.\ Hwa, {\sl Acta Phys.\ Pol.\ } {\bf B27} (1996) 1789,
%
\bibitem{bz} A.\ Bialas, B.\ Ziaja, {\sl Phys.\ Lett.\ }{\bf B378} (1996)
319,
%
\bibitem{jz} R.\ Janik, B.\ Ziaja, {\sl Acta Phys.\ Pol.\ B} (1999) 259,
%
\bibitem{alfa}
D.\ Shertzer, S.\ Lovejoy, {\sl Proc.\ IUATM Symp.\ on Turbulence and Chaotic
Phenomena in Fluids}, Kyoto, Sept.\ 1983 (North-Holland, Amsterdam, 1984);
D.\ Shertzer, S.\ Lovejoy, {\sl Selected Papers from the 4th Symp.\ on
Turbulent Shear Flows}, University of Karlsruhe (1983), ed.\ L.\ J.\ S.\
Bradbury et al.\ (Springer Verlag, 1984)
%
\bibitem{MS87}
C.\ Meneveau, K.\ R.\ Sreenivasan, {\sl Phys.\ Rev.\ Lett.\ }{\bf 59} (1987)
1424,
%
\bibitem{corr} R.\ Peschanski, J.\ Seixas, preprint {\sl CERN-TH-5903-90},
%
\end{thebibliography}
\end{document}